\documentclass[twocolumn,aps,prl,groupedaddress,showpacs]{revtex4-1}
\usepackage{amsfonts}
\usepackage{amsmath}
\usepackage{amssymb}
\usepackage{txfonts}
\usepackage{pxfonts}
\usepackage{graphicx,bm,units,yfonts}
\usepackage[table]{xcolor}

\begin{document}

\title{Quantization of Topological Invariants under Symmetry-Breaking Disorder}
\author{Juntao Song$^{1}$}
\author{Emil Prodan$^{2}$}
\email{prodan@yu.edu}
\affiliation{\ $^{1}$ Department of Physics, Hebei Normal University, Hebei 050024, China
\\$^2$ Department of Physics, Yeshiva University, New York, New York 10016, USA
}

\begin{abstract}
In the strictly periodic setting, the electric polarization of inversion-symmetric solids with and without time-reversal symmetry and the isotropic magneto-electric response function of time-reversal symmetric insulators are known to be topological invariants displaying an exact $\mathbb Z_2$ quantization. This quantization is stabilized by the symmetries. In the present work, we investigate the fate of such symmetry-stabilized topological invariants in the presence of a disorder which breaks the symmetries but restores them on average. Using a rigorous analysis, we conclude that the strict quantization still holds in these conditions. Numerical calculations confirm this prediction.
\end{abstract}


\maketitle

An important issue in the field of symmetry-protected topological phases is the fate of the topological invariants in the presence of bulk and surface disorder which can break the symmetries \cite{FHEA}. For the time-reversal symmetric topological insulators, this issue is related to the contamination of the samples with magnetic impurities and it has been addressed experimentally and theoretically in a number of works \cite{ORV}. While this contamination can be fully controlled in laboratory conditions, this is not the case in the real world conditions. Still, if the contamination is small, the magnetic impurities are in a non-correlated disordered phase and, on average, the time-reversal symmetry is preserved.  Then a question which is is extremely relevant for the practical applications of these materials is if the topological characteristics, such as the extended character of the surface states and the quantization of the bulk topological invariant, are preserved under such ``average" time-reversal symmetry conditions? The first characteristics has been shown in Ref.~\cite{FK} to hold if the disorder is not too strong and here we show that the isotropic magneto-electric response function \cite{QHZ} remains quantized in such disordered regimes, provided a spectral gap is present. The antiferromagnetic topological insulators introduced in Ref.~\cite{MEM} is another class where the topology is stabilized by an average time-reversal symmetry. In the presence of disorder, this class of insulators were recently shown \cite{BEFB} to posses distinct topological phases surrounded by sharp phase boundary which can be detected by transport measurements. Furthermore, the weak topological insulators can be thought as protected by the translational symmetry and disorder definitely breaks this symmetry but the symmetry is restored on average. Refs.~\cite{RKS,MBM} gave evidence that, in certain conditions, this is enough for the surface states of weak topological insulators to avoid Anderson localization. 

When some of the stabilizing symmetries are space symmetries, such as for the crystalline topological insulators \cite{Fu,AF}, the inversion-symmetric \cite{TZMV,HPB}, the reflection-symmetric \cite{CHMR,DPFT} or the spin-orbit free topological insulators \cite{AFGB} in general, these issues are actually of central importance because the space disorder of the atoms is inevitable and impossible to control even in the laboratory conditions. If, for example, one considers the random displacements of the atoms due to finite temperature, there can be no expectations that the disordered potential respects the underlying symmetry of the lattice but one can be sure that, on average, the symmetries are strictly preserved since the thermodynamic state (including the nuclei) is symmetric. Such restoration of the symmetries by averaging also occurs when the disorder is induced by defects which do not destroy the crystalline order completely. Ref.~\cite{FHEA} introduced a $\mathbb Z_2$ topological invariant, which in principle covers all classes of symmetry-protected topological insulators (and more) in such disordered conditions. This topological invariant was put to work for a disordered model with averaged reflection symmetry and a localization-delocalization transition was observed numerically exactly at the point where the invariant changed its value. In the present work, we consider disordered insulators with averaged inversion symmetry and show that the classical electric polarization assumes strict quantized values $0$ or $\frac{1}{2}$, provided a spectral gap is present.

Now, the existing definitions of the symmetry-stabilized topological invariants depend fundamentally on the exactness of the symmetries, hence, at the first sight, it seems impossible to define them for a concrete disorder configuration which, just by itself, breaks the symmetries. However, as noted in Ref.~\cite{FHEA}, when the symmetries, which are preserved only average, are combined with the translations then a certain self-averaging property takes place, enabling one to define {\it exact} topological invariants for such realistic conditions. In the present work, we provide a precise formulation of this self-averaging property within the framework of homogeneous disordered systems \cite{Bel1,Bel2}. While at this moment we cannot make any statements about the boundary states, our findings definitely contributes to the growing body evidence that the symmetry-stabilized topological insulators are more robust than previously thought. 

The paper is organized as follows. In the first section, we describe the homogeneous disordered systems and we  formulate a precise relation between the disorder and symmetry, called here the compatibility condition, which ensures the restoration of the symmetry on average. An explicit yet very generic model for homogeneous disordered solids is presented, and the inversion as well as the time-reversal symmetries are shown to be compatible with the model. The second section discusses the electric polarization of homogeneous disordered systems with inversion symmetry and no time-reversal symmetry. The disorder is assumed to be compatible with the symmetry. The polarization is shown to be an intensive macroscopic function with a self-averaging property (preventing fluctuations from one disorder configuration to another), and which takes only quantized values. The fourth section reports a numerical analysis which confirms the theoretical predictions. The fifth section applies the arguments to the isotropic magneto-electric response function of TRS insulators, leading to similar conclusions. Last section summarizes our conclusions. 

\section{Homogeneous Solids}

Physically, a homogeneous system is defined as an extended system with
translation invariance broken at the microscopic scale but this symmetry-breaking is undetectable at macroscopic scales. In other words, taking micron-size samples from a large piece of a homogeneous material will lead to same
intensive TD parameters and coefficients (of course not exactly the
same for finite pieces, but the differences are well below the
experimental resolution). Formulating this property in a mathematically rigorous way was quite a challenge for the mathematical physics community but these days the concept of a homogeneous solid state system has a very precise meaning and the mathematical framework built around this concept is natural and fruitful for a large class of aperiodic systems \cite{Bel1,Bel2}. According to the precise definition, an aperiodic tight-binding Hamiltonian $H$ over the lattice $\mathbb Z^d$ is homogeneous if $H$ together with its lattice translates $T_{\bm a}H T_{\bm a}^{-1}$, for all $\bm a \in \mathbb Z^d$, define a set which has a compact closure in the strong topology of bounded operators. Here and throughout, $T_{\bm a}$ will denote the lattice translation by $\bm a$. What we are going to present in the following applies strictly to the homogeneous system so defined, but for concreteness we will carry the discussion in the context of disordered lattice models, which are explicitly constructed in Example 1. 

Henceforth, let us consider a generic finite-range disordered lattice model 
\begin{equation}
H_{\bm \omega}=H_0+ V_{\bm \omega}
\end{equation}
defined over the Hilbert space spanned by $| \bm n, \alpha \rangle$, where $\bm n \in \mathbb{Z}^D$ represent the nodes of the lattice indexing the unit cells and $\alpha=1,...,Q$ the orbitals associated with a unit cell. The orbital index includes the spin degree of freedom, but the latter will be separated out when needed. The orbitals can and will be chosen to be real, that is, invariant to complex conjugation: $\mathcal K |\bm n,\alpha \rangle = |\bm n,\alpha \rangle$. The translational invariant piece $ H_0$ of the Hamiltonian is assumed to depend on a set of $N$ parameters $\bm \xi=(\xi_1,\ldots,\xi_N)$. When needed, we write this dependence explicitly as $ H_0(\bm \xi)$ or $ H_{\bm \omega}(\bm \xi)$. We include such dependence because both the electric polarization and the magneto-electric response functions are defined through deformations of the system from a reference configuration. If the reader is more comfortable with an explicit representation of this parameter space, he may think of the vector $\bm \xi$ as the (always finite) collection of hopping amplitudes. 

The random potential $V_{\bm \omega}$ depends on the disorder configuration $\bm \omega$, which is seen here as a point in a disorder configuration space $\Omega$, which is compact and equipped with a probability measure $d \bm \omega$. The system is assumed to be homogeneous, which in this context is assured by the covariant property 
\begin{equation}
T_{\bm a} H_{\bm \omega} T_{\bm a}^{-1}=H_{\mathfrak{t}_{\bm a} \bm \omega}, \quad \forall \bm a \in \mathbb Z^d,
\end{equation}
where $\mathfrak{t}$ represents a homeomorphic action of the group of lattice translations on $\Omega$. The action $\mathfrak t$ is assumed ergodic and probability preserving so that Birkhoff's ergodic theorem \cite{Bir} applies, namely, the following identity holds with probability one in $\omega$ :
\begin{equation}\label{Birkhoff}
\lim_{V \rightarrow \infty} \frac{1}{V}\sum_{\bm a \in V} f(\mathfrak{t}_{\bm a} \bm \omega)=\int_\Omega d\bm \omega' f(\bm \omega').
\end{equation}
It is precisely this identity which ensures the non-fluctuating character of the intensive thermodynamic functions from one disorder configuration to another.

We now formulate a precise condition which automatically leads to the restoration of the symmetry after the disorder average is taken. Consider a generic symmetry operation:
\begin{equation}
\mathcal S |{\bm n},\alpha \rangle = \sum_{\bm n',\alpha'} R_{\bm n \alpha;\bm n' \alpha'}|\bm n',\alpha' \rangle,
\end{equation}
which can be linear or anti-linear. Recall that the orbitals are real, hence the symmetry is fully determined by the coefficients $R_{\bm n \alpha;\bm n' \alpha'}$.  The symmetry generates an action on the parameter space: 
\begin{equation}
\mathcal  S  H_0(\bm \xi) \mathcal  S^{-1}= H_0(\mathcal S \bm \xi),
\end{equation}
and on the disorder configuration space: 
\begin{equation}
\mathcal S  V_{\bm \omega} \mathcal S^{-1}= V_{\mathcal S \bm \omega},
\end{equation} 
for which we use the same notation $\mathcal S$. As a consequence, we have the following covariant property:
\begin{equation}
\mathcal S H_{\bm \omega}(\bm \xi) \mathcal S^{-1}=H_{\mathcal S \bm \omega}(\mathcal S \bm \xi).
\end{equation}
In fact, such covariant property is obeyed by any function of the Hamiltonian:
\begin{equation}
\mathcal S \Phi \big (H_{\bm \omega}(\bm \xi) \big ) \mathcal S^{-1}=\Phi \big (H_{\mathcal S \bm \omega}(\mathcal S \bm \xi) \big).
\end{equation}

\vspace{0.2cm}

\noindent{\bf Definition.} We say that the symmetry is compatible with the disorder if
\begin{equation}\label{Compa}
d (\mathcal S \bm \omega) = d \bm \omega, 
\end{equation}
that is, if the probability measure $d \bm \omega$ is invariant under the action on $\Omega$ induced by the symmetry.
 
\vspace{0.2cm}

Perhaps the reader already noticed that we are merely restating the condition formulated in Ref.~\cite{FHEA} Section IIA, in a more concise mathematical notation. This definition is relevant when we consider the disorder average:
\begin{equation}
\mathcal S \left ( \int_\Omega d \bm \omega \ \Phi \big( H_{\bm \omega}(\bm \xi)\big ) \right ) {\mathcal S}^{-1}=\int_\Omega d \bm \omega \ \Phi \big ( H_{\mathcal S \bm \omega}(\mathcal S \bm \xi) \big ),
\end{equation}
because a change of variable $\mathcal S \bm \omega \rightarrow \bm \omega$ leads to:
\begin{equation}
\ldots =\int_\Omega d (\mathcal S ^{-1} \bm \omega) \ \Phi \big (H_{\bm \omega}(\mathcal S \bm \xi) \big )=\int_\Omega d \bm \omega \ \Phi \big (H_{\bm \omega}(\mathcal S \bm \xi) \big).
\end{equation}
In other words
\begin{equation}
\mathcal S \left ( \int_\Omega d \bm \omega \ \Phi \big( H_{\bm \omega}(\bm \xi)\big ) \right ) {\mathcal S}^{-1} = \int_\Omega d \bm \omega \ \Phi \big (H_{\bm \omega}(\mathcal S \bm \xi) \big).
\end{equation}
In particular, if $\bm \xi$ is a fixed point for the symmetry, $\mathcal S \bm \xi = \bm \xi$, or equivalently if $\mathcal S H_0(\bm \xi) \mathcal S^{-1} = H_0(\bm \xi)$, then
\begin{equation}
\mathcal S \left ( \int_\Omega d \bm \omega \ \Phi \big( H_{\bm \omega}(\bm \xi)\big ) \right ) {\mathcal S}^{-1} = \int_\Omega d \bm \omega \ \Phi \big (H_{\bm \omega}(\bm \xi) \big),
\end{equation}
which shows that Eq.~\ref{Compa} implies the restoration of the symmetry for the averaged quantities.

\vspace{0.2cm}

\noindent{\sl Example 1: Generic homogeneous disordered model.}
\begin{eqnarray}\label{GModel}
H_{\bm \omega}= \sum_{{\bm n},\alpha; \bm m, \beta} \big ( t_{{\bm n}-{\bm m}}^{\alpha \beta}(\bm \xi) +W \omega^{\alpha \beta}_{{\bm n},{\bm m}} \big )|{\bm n},\alpha \rangle  \langle {\bm m},\beta | ,
\end{eqnarray}
where $\omega^{\alpha \beta}_{{\bm n},{\bm m}}$ are independent random entries drawn uniformly from the interval $[-\frac{1}{2},\frac{1}{2}]$. The collection of all random variables $\bm \omega=\{ \omega^{\alpha \beta}_{{\bm n},{\bm m}}\}$ can be viewed as a point in an infinite dimensional configuration space $\Omega$, which is just an infinite product of intervals $[-\frac{1}{2},\frac{1}{2}]$. The result is a compact and metrizable Tychonov space which can be equipped with the product probability measure: 
\begin{equation}
d \bm \omega=\prod_{\bm n,\alpha;\bm m, \beta} d\omega^{\alpha \beta}_{{\bm n},{\bm m}}.
\end{equation} 
There is a natural action of the lattice translations on $\Omega$:
\begin{equation}
(\mathfrak{t}_{\bm a} \bm \omega)^{\alpha \beta}_{{\bm n},{\bm m}}=\omega^{\alpha \beta}_{{\bm n}-{\bm a},{\bm m}-{\bm a}}, ~ \ a\in \mathbb{Z}^{D},
\end{equation}
which leaves $d \bm \omega$ invariant and is known to act ergodically, hence Eq.~\ref{Birkhoff} applies. It is straightforwad to check that the Hamiltonian has indeed the covariant property:
\begin{equation}
T_{\bm a} H_{\bm \omega} T_{\bm a}^{-1} = H_{\mathfrak t_{\bm a} \bm \omega}.
\end{equation}

\vspace{0.2cm}

\noindent{\sl Example 2: Compatibility of the inversion symmetry with the model \ref{GModel}.} This symmetry maps the unit cell $\bm n$ into $-\bm n$ and it can mix the orbitals in the process. However, we can always choose an orbital basis so that no such mixing occurs, and since $\mathcal I^2 = 1$:
\begin{equation}\label{Inversion}
\mathcal I |{\bm n},\alpha \rangle = \sum_{\alpha} \sigma_\alpha |-{\bm n},\alpha \rangle,
\end{equation}
where all $\sigma$'s are signs. The induced action on $\Omega$ is: 
\begin{eqnarray}
(\mathcal I \bm \omega)_{\bm n \bm m}^{\alpha \beta} = \sigma_\alpha \sigma_\beta \omega^{\alpha \beta }_{{-\bm n},{-\bm m}}.
\end{eqnarray}
Now,
\begin{equation}
d \bm \omega=\prod_{\bm n,\alpha;\bm m, \beta} d\omega^{\alpha \beta}_{{\bm n},{\bm m}} = \prod_{\bm n,\bm m; \alpha} d\omega^{\alpha \alpha}_{{\bm n},{\bm m}} \prod_{\bm n,\bm m; \alpha < \beta} d\omega^{\alpha \beta}_{{\bm n},{\bm m}}d\omega^{\beta \alpha}_{{\bm n},{\bm m}}
\end{equation}
and now one can see explicitly that $d (\mathcal I \bm \omega) = d \bm \omega$.

\vspace{0.2cm}

\noindent{\sl Example 3: Compatibility of the time-reversal symmetry with the model \ref{GModel}.} The TRS is defined as the anti-linear map
\begin{equation}\label{Theta}
\Theta |{\bm n},\alpha,\sigma \rangle = -\sigma |\bm n,\alpha,-\sigma \rangle,
\end{equation}
where the spin degree of freedom $\sigma=\pm 1$ (for spin up/down) was separated out. Note that $\Theta^2=-1$. The induced action on $\Omega$ is: 
\begin{eqnarray}
(\Theta \bm \omega)_{\bm n \bm m}^{\alpha,\sigma; \beta,\sigma'} = \sigma \sigma' \omega^{\alpha,-\sigma; \beta,-\sigma'}_{\bm n,\bm m}.
\end{eqnarray}
Then,
\begin{align}
d \bm \omega & =\prod_{\bm n,\alpha,\sigma;\bm m, \beta,\sigma'} d\omega^{\alpha,\sigma; \beta,\sigma'}_{{\bm n},{\bm m}}  \\
&=  \prod_{\bm n,\bm m} d\omega^{\alpha,1; \beta,1}_{{\bm n},{\bm m}}d\omega^{\alpha,-1; \beta,-1}_{{\bm n},{\bm m}}d\omega^{\alpha,-1; \beta,1}_{{\bm n},{\bm m}}d\omega^{\alpha, 1; \beta, -1}_{{\bm n},{\bm m}} \nonumber
\end{align}
and now one can see explicitly that $d (\Theta \bm \omega) = d \bm \omega$.

\section{Electric Polarization of Inversion symmetric insulators} 

\subsection{Generic definition}

By definition \cite{RV,Res}, the change in the electric polarization, as a result of a macroscopic deformation $H_{\bm \omega}(\bm \xi_t)$ of the Hamiltonian, is:
\begin{eqnarray}\label{PDef}
\Delta \bm {\mathcal P}_{\bm \omega} =  \int_0^T  \bm j_{\bm \omega}(t) \, dt,
\end{eqnarray}
where $\bm j_{\bm \omega}(t)$ is the density of the charge-current. The latter is microscopically defined as:
\begin{equation}\label{DCurrent}
\bm j_{\bm \omega}(t)=\lim\limits_{\mathrm{V}\rightarrow \infty}\frac{1}{\mathrm{V}} \sum\limits_{{\bm n} \in \mathrm{V}}\sum_{\alpha=1}^Q\langle \bm n,\alpha|   \rho_{\bm \omega}(t) {\bm J}_{\bm \omega}(t) | \bm n ,\alpha\rangle,
\end{equation}
where $\rho_{\bm \omega}(t)$ is the time-evolved density matrix and 
\begin{equation}
{\bm J}_{\bm \omega}(t) = {\rm i}{\rm e}[H_{\bm \omega}(\bm \xi_t),\bm X]
\end{equation} 
is the current operator. Here, $\bm X$ denotes the position operator and, for convenience, the electron charge ${\rm e}$ will be set to unity in the following. If the deformation starts from the thermodynamic equilibrium state, then the time-evolved density matrix is 
\begin{equation}
\rho_{\bm \omega}(t)=U_t \Phi_{\mathrm{FD}}\big (H_{\bm \omega}(\bm \xi_0)\big) U_t^{-1},
\end{equation}
 with $U_t$ being the unitary time evolution generated by $H_{\bm \omega}(\bm \xi_t)$ and $\Phi_{\mathrm{FD}}$ the Fermi-Dirac distribution. One should not confuse $\rho_{\bm \omega}(t)$ and $\Phi_{\mathrm{FD}}\big (H_{\bm \omega}(\bm \xi_t)\big)$, because the time-evolved density matrix is no longer given by the Gibbs state.
 
 \subsection{Self-averaging property}
 
 Now note that the fundamental formula in Eq.~\ref{PDef} involves the density of the current rather than the current itself, and this is why the trace per volume appears in Eq.~\ref{DCurrent}. This is consistent with the fact that polarization is an intensive macroscopic function. From a technical point of view, this is an important observation because we can use Birkhoff's ergodic theorem \cite{Bir} to demonstrate that $\Delta \bm {\mathcal P}_{\bm \omega}$ is, with probability one, independent of the disorder configuration $\bm \omega$. Indeed, note that the operator inside the brackets of Eq.~\ref{DCurrent}, called $F_{\bm \omega}$ in the lines below, is covariant:  
 \begin{equation}
 T_{\bm a} F_{\bm \omega} T_{\bm a}^{-1} = F_{t_a \bm \omega}.
 \end{equation}
  Then:
\begin{equation}\label{TraceCell}
\tfrac{1}{\mathrm{V}} \sum\limits_{{\bm n} \in \mathrm{V}}\langle {\bm n}| F_{\bm \omega} | \bm n \rangle
= \tfrac{1}{\mathrm{V}} \sum\limits_{{\bm n} \in \mathrm{V}} \langle 0| F_{t_{\bm n}^{-1}\bm \omega} | 0 \rangle 
    =  \int\limits_\Omega d \bm \omega \; \langle 0 | F_{\bm \omega} | 0  \rangle 
\end{equation}
in the limit $\mathrm{V}\rightarrow \infty$. The conclusion is that $\Delta \mathcal P _{\bm \omega}$ is self-averaging and its macroscopic value comes as an average over the disorder configuration space. As such, we can drop the subscript $\bm \omega$ in $\Delta \bm {\mathcal P}_{\bm \omega}$ from now on.

\subsection{Schulz-Baldes-Teufel formula}

We consider now infinitely slow deformations of the Hamiltonian, which are better visualized as paths $\gamma$ in the parameter space, parametrized as $\{\bm \xi_s\}_{s\in [0,1]}$. By employing the adiabatic theorem, Schulz-Baldes and Teufel showed in Ref.~\cite{ST} that in the extreme adiabatic limit and when the temperature goes to zero:
\begin{equation}\label{SBT0}
\Delta \bm {\mathcal P} (\gamma) = \int_0^1 ds \; \mathcal T \big ( P_{\bm \omega}(\bm \xi_s) {\rm i} \big [ \partial_s  P_{\bm \omega}(\bm \xi_s),\mathrm{i} [\bm X, P_{\bm \omega}(\bm \xi_s)] \big ]  \big ),
\end{equation}
where $\mathcal T$ denotes the trace per volume:
\begin{equation}
\mathcal T( \ldots ) = \lim_{V \rightarrow \infty} \tfrac{1}{V} \sum_{\bm n \in V} \sum_{\alpha =1}^Q \langle \bm n,\alpha | \ldots |\bm n,\alpha \rangle,
\end{equation}
and
\begin{equation}
P_{\bm \omega}(\bm \xi)=\chi_{[-\infty,\epsilon_F]} \big ( H_{\bm \omega}(\bm \xi) \big )
\end{equation}
 is the Fermi projection onto the occupied states at coordinate $\xi$. Above, it is assumed that the spectral gap of $H_{\bm \omega}(\bm \xi_s)$ remains open for all $s \in [0,1]$ and that the Fermi level $\epsilon_F$ is inside this gap. We will keep the ${\rm i} = \sqrt{-1}$ in front of the commutators in order to make them self-adjoint. Now, using Birkhoff's theorem as before, we can write, equivalently:
\begin{align}\label{SBT}
\Delta \bm {\mathcal P} (\gamma) = & \int_0^1 ds \int_\Omega d \bm \omega \ \sum_{\alpha=1}^Q \medskip \\
& \big \langle 0,\alpha \big | P_{\bm \omega}(\bm \xi_s) {\rm i} \big [ \partial_s  P_{\bm \omega}(\bm \xi_s),\mathrm{i} [\bm X, P_{\bm \omega}(\bm \xi_s)] \big ]  \big | 0,\alpha \big \rangle. \nonumber
\end{align}
This is Schulz-Baldes-Teufel formula, which can be regarded as the disordered version of the King-Smith-Vanderbilt formula for the static spontaneous orbital polarization \cite{KV}.

\subsection{Quantization}

\begin{figure}[!ht]
{\includegraphics[width=0.8\columnwidth]{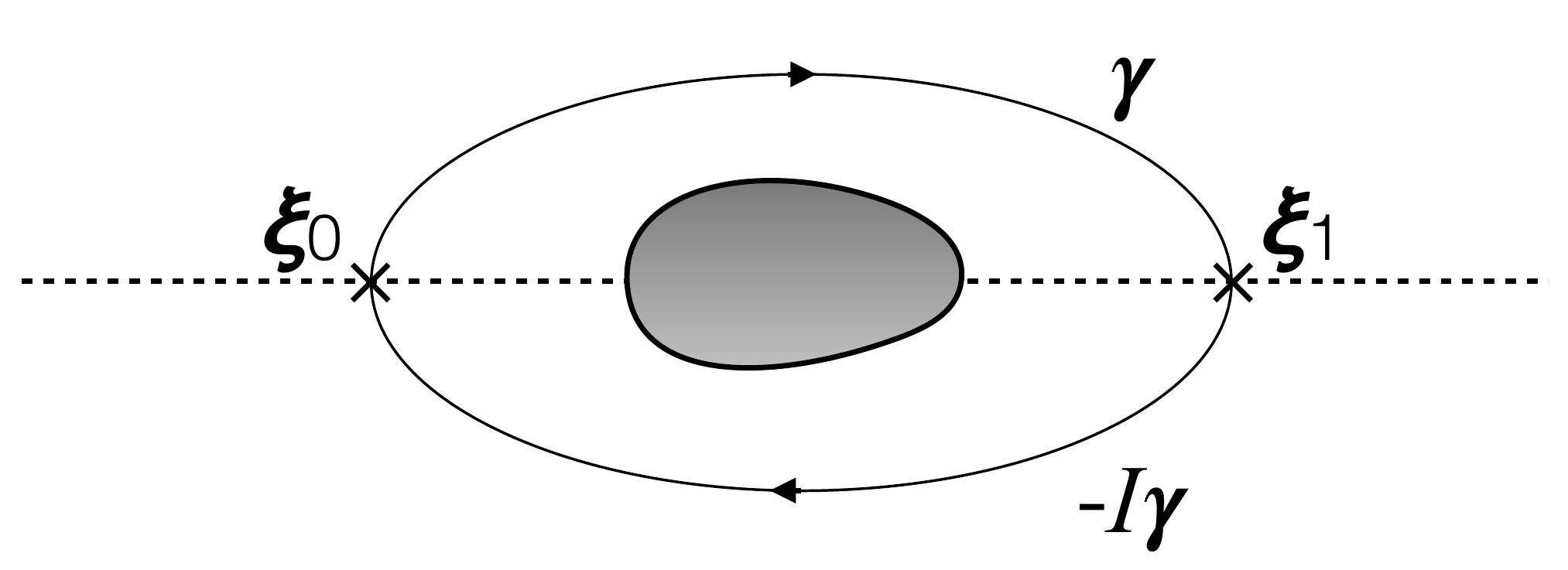}}
\caption{The direct path $\gamma$ (upper semi-plane) and the inverted path $-\mathcal I \gamma$ (lower semi-plane) in the parameter space $\xi$. The horizontal axis represents the manifold which left invariant by the symmetry. The interesting cases occur when the spectral gap closes along this manifold, which is schematically shown by the gray region.}
\label{Fig0}
\end{figure}

The proof of quantization proceeds in two steps. First, based on Schulz-Baldes-Teufel formula Eq.~\ref{SBT}, we can demonstrate that the change in the electric polarization along the inverted path ${\mathcal{I}} \gamma$ (see Fig.~\ref{Fig0}) is: 
\begin{equation}\label{A2}
\Delta \bm {\mathcal P} ( {\mathcal{I}}\gamma)=-\Delta \bm {\mathcal P} (\gamma).
\end{equation}
This equality is remarkable because we are computing the polarization of a system in an arbitrary disorder configuration which breaks the inversion symmetry. For the periodic case with strict inversion symmetry, this property is well known \cite{TZMV,HPB}. The keys to its proof are the self-averaging property, the compatibility between the inversion symmetry and disorder, and the behavior of the position operator under inversion, $\mathcal I \bm X \mathcal I^{-1} = -\bm X$. The proof proceeds as follows:
\begin{align}
\Delta \bm {\mathcal P} ({\mathcal I} \gamma) & =  \int_0^1 ds \int_\Omega d \bm \omega \ \sum_{\alpha=1}^Q  \\
 & \big \langle 0,\alpha \big | P_{\bm \omega}({\mathcal I} \bm \xi_s) {\rm i}\big [ \partial_s P_{\bm \omega}({\mathcal I} \bm \xi_s), \mathrm{i}[\bm X, P_{\mathcal \bm \omega}({\mathcal I}\bm \xi_s)] \big ] \big  | 0,\alpha \big \rangle, \nonumber
\end{align}
and observe that 
\begin{equation}
P_{\bm \omega}({\mathcal I} \bm \xi_s)=\mathcal I P_{{\mathcal I}^{-1}\bm \omega}(\bm \xi_s) \mathcal I^{-1}.
\end{equation} 
Then a change of variable $ \bm \omega \rightarrow \mathcal{I} \bm \omega$ and cancelations of terms like $\mathcal I \mathcal I^{-1}$, lead us to:
\begin{align}
\Delta \bm {\mathcal P} ({\cal I} \gamma) & =\int_0^1 ds \int_\Omega d (\mathcal I \bm \omega) \ \sum_{\alpha=1}^Q \\
& \big \langle 0,\alpha \big | \mathcal I P_{\bm \omega}(\bm \xi_s) {\rm i}\big [ \partial_s P_{\bm \omega}(\bm \xi_s), \mathrm{i}[-\bm X, P_{\cal \bm \omega}(\bm \xi_s)] \big ] \mathcal I^{-1} \big | 0,\alpha \big \rangle. \nonumber
\end{align}
The probability measure is invariant, $d (\mathcal{I} \bm \omega) = d{\bm \omega }$. Also, the $0$-site is left invariant by $\mathcal I$ and, since we are tracing over the orbital degrees of freedom, the action of the remaining $\mathcal I$ operators have no effect and can be removed. Then Eq.~\ref{A2} follows.

The second step of the proof consists of the following argument. Assume that the initial and final Hamiltonians are inversion symmetric, that is, $\bm \xi_0$ and $\bm \xi_1$ are fixed points of $\mathcal I$, $\mathcal I\bm \xi_0 = \bm \xi_0$ and $\mathcal I\bm \xi_1=\bm \xi_1$. Then any path $\gamma$ joining the two Hamiltonians can be closed into a loop by augmenting $-\mathcal{I}\gamma$, where the minus sign imply that the path is walked in reverse (see Fig.~1). Note that the argument doesn't work if the end points of $\gamma$ are not fixed points for $\mathcal I$. Given Eq.~\ref{A2}, this has the following effect:
\begin{equation}
\Delta \bm {\mathcal P} (\gamma)=\nicefrac{1}{2}\big(\Delta \bm {\mathcal P} (\gamma)+\Delta \bm {\mathcal P} (-{\cal I} \gamma)\big)=\nicefrac{1}{2}\Delta \bm {\mathcal P} (\gamma -{\cal I} \gamma ).
\end{equation}
But as already noted in \cite{NL,ST}, the change in the polarization along a closed loop leads to the non-commutative 1st Chern number defined in Ref.~\cite{BES}. The conclusion is:
\begin{equation}\label{Chern}
\Delta \mathcal P_j (\gamma)=\nicefrac{1}{2} \; C_1[(\gamma -{\cal I} \gamma )\times \mathcal S_j], \ j=1,\ldots,D,
\end{equation}
where on the righthand side is the Chern number of the Fermi projection over the manifold $(\gamma -{\cal I} \gamma )$ times the section of the non-commutative Brillouin torus along the $j$th direction:
\begin{equation}\label{Chern1}
C_1 = \int\limits_{\gamma -\mathcal I \gamma} d\bm \xi \; \mathcal T \big ( P_{\bm \omega}(\bm \xi) {\rm i} \big [ \partial_\xi  P_{\bm \omega}(\bm \xi),\mathrm{i} [X_j, P_{\bm \omega}(\bm \xi)] \big ]  \big ).
\end{equation}
The usual constant in front is absent here because one derivative is taken in the `$k$'-space and the other in the real-space representations. As longs as $P_{\bm \omega}(\bm \xi)$ is smooth along the loop and its matrix elements $\langle \bm n|P_{\bm \omega}(\bm \xi)|\bm m\rangle$ decay sufficiently fast, all these Chern numbers are integers for dimension $D=1$ or $2$.  In $D=3$ these Chern numbers are weak topological invariants and they remain integer only if magnetic fields are not present. All these conditions are met if the spectral gap at the Fermi level remains open along the loop. 

In conclusion, when insulators are deformed between fixed points of the inversion symmetry operator, the change in the electric polarization is quantized in units of $\tfrac{1}{2}$, even thought the symmetry exists only for disorder averages.

\begin{figure*}[!ht]
{\includegraphics[width=1.8\columnwidth]{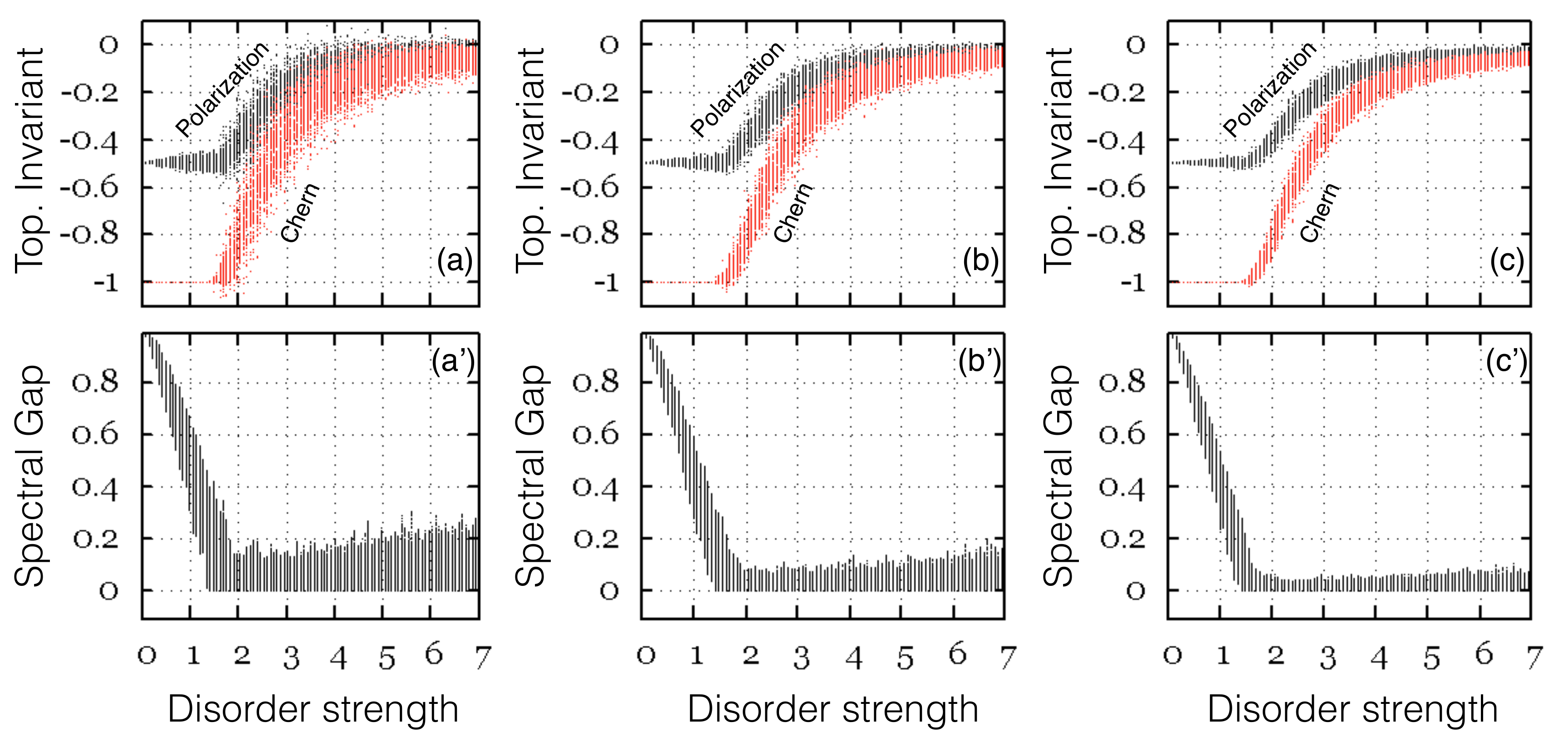}}
\caption{ (Color online) Top row: The variation of the electric polarization when the Rice-Mele model is adiabatically deformed along a semi-circle (black dots) and along a full circle (red dots) in the two-parameter space. In the latter case, the variation is equal to a Chern number, hence the label. The calculations are performed in the presence of a disordered potential which breaks the inversion-symmetry but restores it on the average. The disorder strength was increased from 0 to 7. Each dot corresponds to one disorder configuration and 100 disorder configurations were considered for each disordered strength. The average and statistical variance are not shown on purpose (see text). Bottom rows: The spectral gap of the Hamiltonian at half-filling. A dot represents the spectral gap at one disorder configuration and one position along the adiabatic deformation path. The three columns correspond to the different system sizes: $L = 100$ (a), 200 (b) and 400 (c).}
\end{figure*}

\subsection{Numerical confirmation} 

We consider the 1-dimensional Rice-Mele model \cite{RM}:
\begin{align}\label{RiceMele}
H_{\bm \omega}(\bm \xi) = &\tfrac{1}{2}\sum_{n \in \mathbb Z} \big(1+{(-1)}^{n} \xi_1\big) \big (|n\rangle \langle n+1|+|n+1\rangle \langle n| \big) \\
& +\tfrac{1}{2}\sum_{n \in \mathbb Z} \big((-1)^n\xi_2+W\omega_n\big)\; |n\rangle \langle n|,\nonumber
\end{align}
with onsite disorder. As detailed in \cite{XCN}, this model was introduced originally by Rice and Mele to study solitons in conducting polymers but it was subsequently used in other contexts too. It is a particular form of the generic model Eq.~\ref{GModel}, hence everything we stated about Eq.~\ref{GModel} continue to hold here. In the absence of disorder, the Rice-Mele model is gapped if half-filling is assumed, except at $\bm \xi =0$, and it is inversion-symmetric whenever $\xi_2=0$. Also, it is well known \cite{XCN} that adiabatic deformations around closed loops surrounding the origin leads to a quantized polarization change $\Delta \mathcal P = \pm 1$ (or $C_1= \pm 1$).

The model has two states per unit cell and it can be brought to the form written in Eq.~\ref{GModel} by defining 
\begin{equation}
|2n\rangle =|n,1\rangle, \quad |2n+1 \rangle = |n,-1\rangle,
\end{equation}
 in which case the Hamiltonian takes the form:
\begin{align}
&H_{\bm \omega}(\bm \xi) = \tfrac{1}{2}\sum_{n \in \mathbb Z} \sum_{\alpha=\pm 1}\big(1+\alpha \xi_1\big) \\
& \times \big (  \big |n,\alpha \big \rangle \big \langle n +\tfrac{1}{2}(1-\alpha),-\alpha \big| + \big | n +\tfrac{1}{2}(1-\alpha),-\alpha \big \rangle \big \langle n,\alpha \big | \big ) \nonumber  \\
& +\tfrac{1}{2}\sum_{n \in \mathbb Z} \sum_{\alpha = \pm 1}\big(\alpha \xi_2+W\omega_{n,\alpha}\big)\; |n,\alpha\rangle \langle n,\alpha|.\nonumber
\end{align}

If in the original rendering of Eq.~\ref{RiceMele} we choose the inversion point to be between the $0$-th and first site, then the inversion operation takes form: 
\begin{equation}
\mathcal I |n,\alpha \rangle \rightarrow |-n,-\alpha \rangle.
\end{equation}
It can be set in the diagonal form of Eq.~\ref{Inversion} if we choose to work with the states $\tfrac{1}{\sqrt{2}}(|n,1\rangle \pm n,-1 \rangle)$. However, we will not do that here. A direct computation will show that:
\begin{equation}
\mathcal I H_{\bm \omega} \mathcal I^{-1} = H_{\mathcal I \bm \omega}(\mathcal I \bm \xi),
\end{equation}
with 
\begin{equation}
\mathcal I (\xi_1,\xi_2) = (\xi_1,-\xi_2), \quad (\mathcal I \bm \omega)_{n,\alpha} = \omega_{-n,-\alpha}.
\end{equation}
Let us stress again that the onsite disorder breaks the inversion symmetry because we impose no correlation between $\omega_{n,\alpha}$ and $\omega_{-n,-\alpha}$. However, the probability measure $\prod_{n,\alpha} d \omega_{n,\alpha}$ is easily seen to be invariant under $\mathcal I$.  

We now choose the path $\gamma$ to be the semi-circle:  
\begin{equation}
\gamma =0.1 \times \{\cos s, \sin s \}_{s\in[0,\pi]},
\end{equation}
which connects two fixed points of the inversion symmetry. By augmenting with $-\mathcal I \gamma$ we obtain the full circle, $s\in [0,2\pi]$. The numerical calculations are performed at finite size $n\in \{0,\ldots,L\}$ and with periodic boundary conditions. Formulas from Eqs.~\ref{SBT0} and \ref{Chern1} are used for all the results reported here. The commutator with the position operator is implemented using the strategy developed in Ref.~\cite{Pro}. The path integral is discretized using 100 points for $\gamma$ and $200$ points for $\gamma-\mathcal I \gamma$. The derivative $\partial_\xi$ with respect to $\bm \xi$ along the path is computed using the five-point stencil finite-difference approximation. The disorder strength $W$ was increased gradually and the calculation was repeated 100 times with updated random potential for each $W$. 

The results are reported in Fig.~2, where each dot seen in these plots represents a single disorder configuration. In other words, {\it no disorder average} has been performed on the data. In the first row we show the results for $\Delta \mathcal P(\gamma)$ and Chern number or $\Delta \mathcal P(\gamma-\mathcal I \gamma)$. The first thing to notice is the clustering of the data for $\Delta \mathcal P(\gamma)$ around the predicted quantized value of $-\tfrac{1}{2}$, at disorder strengths lower than a critical value $W_c$. The fluctuations around the quantized values can be attributed to the finite system sizes because the fluctuations can be seen to diminish as the size of the system gets larger. This confirms our theoretical prediction that $\Delta \mathcal P(\gamma)$ is self-averaging hence non-fluctuating in the thermodynamic limit, and that, at least for small $W$'s, it takes quantized values in the unit of $\tfrac{1}{2}$. 

The Chern number shows a very good quantization, with virtually no fluctuations from one disorder configuration to another below the same $W_c$. The reason for this difference is that the quantization of the Chern number does not require the restoring of the symmetry. It only requires that the exponential decay rate of the Fermi projection be large compared to $1/L$. The Fermi projection also needs to be smooth of $\bm \xi$ along the loop, but this is automatically the case if the spectral gap remains open (which is the case for small $W$'s).  If, for example, we average $\Delta \mathcal P(\gamma)$ over just two disorder configurations which are mirrored by inversion symmetry, then the quantization of $\Delta \mathcal P(\gamma)$ will be as good as that of the Chern number (this can be shown exactly).  While this could be a better numerical method to compute $\Delta \mathcal P(\gamma)$, it is irrelevant for the present discussion because here we want to demonstrate a principle, namely, that the quantization of $\Delta \mathcal P(\gamma)$ occurs even for a single disorder configuration $\bm \omega$, provided the system size is large enough. So, what is actually happening in this latter case? For this, let us note that the formula for $\Delta \mathcal P(\gamma)$ remains invariant if the disorder is translated. Now, due to the ergodicity of the translations, when we translate the disorder we start exploring the disorder configuration space $\Omega$, and if the size of the system is large enough, at some point a translation will bring us close to a new disorder configuration which looks in many respects like the inverted disorder $\mathcal I \omega$. In other words, if the system size is large enough, we don't need to average over two mirrored disorder configurations because this already happens due to the ergodicity of the system. 

The last issue we need to address is the value of the critical $W_c$. We know already that $W_c$ should be larger or equal than the value of $W$ where the spectral gap of $H_{\bm \omega}(\bm \xi)$ closes somewhere along the loop. Can $W_c$ be strictly larger than this value that we mentioned? To answer this question, we plotted in the second row of Fig.~2 the spectral gaps of $H_{\bm \omega}(\bm \xi)$ at half-filling, as $\bm \xi$ was varied along $\gamma$ and the disorder configurations were updated. These many data were then collapsed and shown for increasing values of $W$. Whenever a dot in the second row of Fig.~2 touches the horizontal axis, the spectral gap of  $H_{\bm \omega}(\bm \xi)$ closed for some $\bm \xi$ along the loop and some disorder configuration. Note that the fluctuations of the gap due to disorder die out in the thermodynamic limit. By comparing the first and second rows of Fig.~2, we can conclude with confidence that the deviations from the quantized values in the top row of Fig.~2 occur exactly at the value of $W$ where the spectral gap closes. The answer to our question is no, and the reason is because the Fermi projection fails to be smooth of $\bm \xi$ beyond that point. 

\section{Magneto-electric response of TRS insulators}

Consider now an insulator $H_{\bm \omega} (\bm \xi)$ in dimension $D=3$. In this section we investigate the isotropic part of the magneto-electric response function:
\begin{equation}
\alpha = \tfrac{1}{3} \sum_{j=1}^3  \frac{\partial \, \mathcal P_j}{\partial B_j} .
\end{equation}
If TRS is considered, then the partial derivatives with respect to the magnetic field are taken at $\bm B=0$. Since the arguments are repetitive, we will expedite the exposition. 

The main tool of our analysis is the formula derived in Ref.~\cite{LP} for the change $\Delta \alpha$ during a deformation of the system along a path $\gamma =\{\bm \xi_s \}_{s \in [0,1]}$ in the parameter space:
\begin{equation}
\Delta \alpha(\gamma)  = \tfrac{1}{2} \epsilon^{i_1 \ldots \i_4} \int_0^1 ds \, \mathcal T\big ( P_{\bm \omega}(\bm \xi_s) \partial_{i_1} P_{\bm \omega}(\bm \xi_s) \ldots \partial_{i_4} P_{\bm \omega}(\bm \xi_s) \big ) \nonumber
\end{equation}
where $\mathcal T$ denotes the trace per volume, $\partial_j$ is a shorthand for:
\begin{equation}
\partial_j P_{\bm \omega}(\bm \xi_s) = i [X_j,P_{\bm \omega}(\bm \xi_s)], \quad \mbox{for} \ j=1,2,3,
\end{equation}
and $\partial_4 P_{\bm \omega}(\bm \xi_s) = \partial_s P_{\bm \omega}(\bm \xi_s)$. Also, $\epsilon^{i_1 \ldots \i_4}$ is the antisymmetric tensor and summation over repeating indices is assumed. Since the operator inside the trace per volume is covariant, we can apply Birkhoff's ergodic theorem \cite{Bir} to equivalently write
\begin{align}
\Delta \alpha (\gamma) = & \tfrac{1}{2} \epsilon^{i_1 \ldots \i_4} \int_0^1 ds \, \int_\Omega d\bm \omega \\
&\sum_{\alpha} \big \langle \bm 0,\alpha \big | P_{\bm \omega}(\bm \xi_s) \partial_{i_1} P_{\bm \omega}(\bm \xi_s) \ldots \partial_{i_4} P_{\bm \omega}(\bm \xi_s) \big | \bm 0,\alpha \big \rangle. \nonumber
\end{align}
This shows at once the self-averaging property of the magneto-electric response function.

Next, we show that if the deformation occurs along the TRS mirrored path $\Theta \gamma$, then:
\begin{equation}
\Delta \alpha(\Theta \gamma) = - \Delta \alpha(\gamma).
\end{equation}
This property is well known for periodic \cite{QHZ} and disordered \cite{LP} TRS insulators, but here we compute the magneto-electric effect for a system in a disorder configuration which breaks the TRS. The proof proceed the same way as for polarization, using the self-averaging property, the compatibility between TRS and disorder, together with the behavior of the derivatives under TRS: $\Theta \partial_j \Theta^{-1} = -\partial_j$, for $j=1,2,3,$ and $\Theta \partial_s \Theta^{-1} = \partial_s$. The proof goes as follows:
\begin{align}
\Delta \alpha (\Theta\gamma) = & \tfrac{1}{2} \epsilon^{i_1 \ldots \i_4} \int_0^1 ds \, \int_\Omega d\bm \omega \\
&\sum_{\alpha} \big \langle \bm 0,\alpha \big | P_{\bm \omega}(\Theta \bm \xi_s) \partial_{i_1} P_{\bm \omega}(\Theta \bm \xi_s) \ldots \partial_{i_4} P_{\bm \omega}(\Theta \bm \xi_s) \big | \bm 0,\alpha \big \rangle. \nonumber
\end{align}
Recall that 
\begin{equation}
P_{\bm \omega}(\Theta \bm \xi_s) = \Theta P_{\Theta^{-1}\bm \omega}( \bm \xi_s) \Theta^{-1}.
\end{equation} 
Then, after a change of variable $\bm \omega \rightarrow \Theta \bm \omega$:
\begin{align}
\ldots = & -\tfrac{1}{2} \epsilon^{i_1 \ldots \i_4} \int_0^1 ds \, \int_\Omega d(\Theta \bm \omega) \sum_{\alpha} \\
& \big \langle \bm 0,\alpha \big | \Theta P_{\bm \omega}(\bm \xi_s) \partial_{i_1}  P_{\bm \omega}( \bm \xi_s) \ldots \partial_{i_4}  P_{\bm \omega}( \bm \xi_s) \Theta^{-1} \big | \bm 0,\alpha \big \rangle, \nonumber
\end{align}
Then the statement follows from the explicit action of $\Theta$ given in Eq.~\ref{Theta} and the compatibility between TRS and disorder, $d(\Theta \bm \omega) = d\bm \omega$.

Lastly, given a path $\gamma$ between two TRS fixed points, $\bm \xi_0 = \Theta \bm \xi_0$ and $\bm \xi_1 = \Theta \bm \xi_1$, one can close this path into a loop by augmenting with its TRS mirrored image $- \Theta \gamma$, taken with opposite orientation. Note that this argument will not work if the end-points of $\gamma$ are not TRS fixed points. Then:
\begin{equation} 
\Delta \alpha(\gamma) = \nicefrac{1}{2}\big ( \Delta \alpha(\gamma) + \Delta \alpha(-\Theta \gamma) \big ) = \nicefrac{1}{2} \Delta \alpha(\gamma -\Theta \gamma) .
\end{equation}
For the periodic TRS case, it is well known \cite{QHZ} that the variation of the magneto-electric response function along a closed loop is equal to a second Chern number. As demonstrated in Ref~\cite{LP}, this remains true in the disordered case, in which case the connection is with the non-commutative second Chern number introduced in Ref.~\cite{PLB}. More precisely:
 \begin{equation}
 \Delta \alpha(\gamma -\Theta \gamma) = C_2 \big [(\gamma - \Theta \gamma) \times \widetilde{\mathbb T}^3 \big ],
 \end{equation}
 where on the right we have the second Chern number of the Fermi projection $P_{\bm \omega}(\bm \xi)$ over the manifold $(\gamma - \Theta \gamma)$ times the 3-dimensional non-commutative torus:
 \begin{equation}
C_2  = \epsilon^{i_1 \ldots \i_4} \int\limits_{\gamma - \Theta \gamma} d \bm \xi \; \mathcal T\big ( P_{\bm \omega}(\bm \xi) \partial_{i_1} P_{\bm \omega}(\bm \xi) \ldots \partial_{i_4} P_{\bm \omega}(\bm \xi) \big ).
\end{equation}
This is a strong topological invariant which is known \cite{PLB} to take only integer values. Note that the constant in front differs from the usual constant because some of the derivatives are taken in $k$-space and some in real-space.
 
 In conclusion, when 3-dimensional insulators are deformed between fixed points of the time-reversal operation, the change in the isotropic magneto-electric response function is quantized in units of $\tfrac{1}{2}$, even thought TRS applies only for disorder averages. Recall that the existence of a spectral is required by our argument. The numerical simulations of $\Delta \alpha$ performed for the work Ref.~\cite{Leu} indicate that, as in the previous case, the quantization does not survive beyond the spectral gap closing. 

\section{Conclusions}

The present work dealt exclusively with the bulk invariants, while the invariants formulated in Ref.~\cite{FHEA} can be regarded as boundary invariants since they are computed from the boundary states. We can already foresee a connection between these invariants, which we would like to sketch briefly. The bulk-boundary principle developed in Ref~\cite{SKR} provides an equality between the bulk first Chern number and a certain spectral flow of the boundary states. The latter has a similar self-averaging property as the first Chern number. When applied to the variation of the electric polarization $\Delta P(\gamma - \mathcal I \gamma)$ of a 1-dimensional system with an edge, this bulk-boundary principle seems to lead precisely to the counting of the spectral features of the edge states performed in Ref.~\cite{FHEA}. It will definitely be interesting to make this connection more precise and see it in action for concrete models.

The principle described in the present work seems to apply to any symmetry-stabilized topological invariant which can be formulated in a real space representation. Unfortunately, there are important instances where a real space representation is not yet available, such as the Kane-Mele $\mathbb Z_2$ invariant \cite{KM} or the bulk topological invariants for point-symmetry stabilized topological insulators \cite{FGB}. For this reason, we have nothing to say about these invariants at this moment, but we hope our findings will spur a renewed effort in this direction. Nevertheless, the strategy does apply to the Loring-Hastings invariants \cite{HL1,HL2}, or to the spin-Chern numbers \cite{SWSH,Prod}.

\section*{ACKNOWLEDGMENTS}
This work was supported by the U.S. NSF grant DMR-1056168 and NSFC grant No.~11204065 and NSFHPC grant No.~A2013205168.

\end{document}